\documentclass[twocolumn]{aastex62}

\usepackage{amsmath}
\usepackage{amssymb}
\usepackage{natbib}
\usepackage{latexsym}

\usepackage{graphicx}
\usepackage{dcolumn}
\usepackage{bm}

\usepackage[T1]{fontenc}
\usepackage[utf8]{inputenc}

\usepackage{xcolor}
\usepackage{xspace}

\usepackage{lineno}

\newcommand{\nn}{ \nonumber }

\usepackage{scalerel}

\newcommand{\beq}{\begin{equation}}
\newcommand{\eeq}{\end{equation}}
\newcommand{\beqa}{\begin{eqnarray}}
\newcommand{\eeqa}{\end{eqnarray}}

\newcommand{\change}[1]{\textcolor{black}{ #1 }}

\usepackage[normalem]{ulem}

\begin{document}


\title{Density-Shear Baryon Acoustic Oscillation as a Cosmological Consistency Check }

\author{Kwan Chuen Chan}
\email{chankc@mail.sysu.edu.cn}
\affiliation{ School of Physics and Astronomy, Sun Yat-Sen University, 2 Daxue Road, Tangjia, Zhuhai 519082, China }
\affiliation{ CSST Science Center for the Guangdong-Hongkong-Macau Greater Bay Area, SYSU, Zhuhai 519082, China }

\author{Yin Li}

\affiliation{ Department of Strategic \& Advanced Interdisciplinary Research, Peng Cheng Laboratory, Shenzhen 518066, China }

\author{Jamie McCullough}
\affiliation{ Department of Astrophysical Sciences, Princeton University, Peyton Hall, Princeton, NJ 08544, USA } 

\date{\today}

\begin{abstract}

  Tensions often arise between different datasets in cosmology, and consistency tests can serve as a powerful tool for diagnosing potential issues. Density-shear Baryon Acoustic Oscillation (GI BAO) is the imprint of the BAO feature on the shear field induced by the large-scale tidal field. We highlight that GI BAO can provide a robust consistency check for the density BAO, shear measurement, and alignment model. Failure of this check hints at systematics in any of these parts.  As an illustration, we present the first GI BAO measurement on photometric data using the Dark Energy Survey Year 3 dataset, achieving a detection significance of $0.86 \sigma$.   We find the GI BAO constraint on the BAO scale dilation parameter $\alpha $  to be $ 0.966 \pm 0.252 $ (1$\sigma$), in good agreement with the density BAO constraint, $ 0.966 \pm 0.037 $, thereby validating the density BAO, shear measurement, and the linear alignment model. Furthermore, we argue that combining the density BAO with GI BAO yields results that are more resilient to systematic effects. Thanks to the massive data volumes of stage IV surveys, GI BAO will play an even more prominent role as a consistency check.

\end{abstract}



\section{Introduction }
\label{sec:intro}

As cosmological measurements become more precise, tensions or inconsistencies with the standard model may emerge, hinting at uncorrected systematics or, more intriguingly, new physics \citep{PerivolaropoulosSkara_2022,DiValentino_etal2025}.  Today, the most well-known tension is the Hubble tension, i.e.,~the Hubble parameter inferred from the CMB differs from the local SNe measurements by more than 5$\sigma$ \citep{Verde_etal2024}. The $S_8$ tension is another notable  example, in which the $S_8$ value inferred from CMB is higher than the lensing constraints.  While some analyses point to settle this tension \citep{McCullough_etal2024,Bigwood_etal2024, Piccirilli_etal2025, Wright_etal2025}, yet it persists in others \citep{Porredon_etal2025,DES_Y63x2pt2026}. Consistency checks across datasets help address these tensions.

Baryon Acoustic Oscillation (BAO) is the imprint of the sound horizon feature in the large-scale structure due to the acoustic oscillations in the primordial plasma, and it is generally regarded as a standard ruler in cosmology \citep{SunyaevZeldovich1970,PeeblesYu1970}. This feature provides a powerful cosmological probe, enabling precise measurements of the universe's expansion history and geometry \citep{Weinberg_etal2013}. BAO has been measured across cosmic time by numerous survey analyses \citep{Eisenstein_etal2005,Cole_etal2005, Gaztanaga:2008xz, Beutler_etal2011, Anderson_BOSS2012, Kazin_etal2014, Alam_etal2017, DES_Y1BAO2019, eBOSS:2020yzd, DES_Y3BAO2022,Chan_DESY3BAO2022, DES_Y6BAO2024, DESI_BAO1_2024,DESI_BAO2_2025}. In particular, the DESI BAO measurements have raised the possibility of dynamical dark energy evolution \citep{DESI_BAO1_2024,DESI_BAO2_2025}.

In this work, we point out that the density-shear BAO (GI BAO) \citep{ChisariDvorkin2013,OkumuraTaruya_2020,Xu:2023vrl} can serve as a powerful consistency check in cosmology. GI BAO is the imprint of the BAO signals on the shear field due to large-scale tidal effects. \citet{vanDompseler_etal2023} show, via a simple physical model, that the density excess producing the BAO peak in the density correlation also creates a trough in the density-shear correlation.  An important consequence is that if we measure the density BAO and GI BAO in the {\it same} galaxy sample, we should get consistent results. In particular, there is no sample variance between them because the GI BAO just measures the same underlying modes as the (scalar) density one but is expressed as the spin-2 field. This enables us to check the consistency of the density BAO measurement, shear measurement, and alignment model. Discrepancy between them hints at the existence of untreated systematics in any of these parts.

As an example, we measure the GI BAO on a photometric data sample from the Dark Energy Survey (DES) Y3 and demonstrate its consistency with the density BAO measurement. DES has presented the transverse (density) BAO measurement in  Y1 \citep{DES_Y1BAO2019}, Y3 \citep{DES_Y3BAO2022}, and Y6  \citep{DES_Y6BAO2024} analyses.  DES Y3 BAO analysis yields a 2.7\% constraint on the transverse BAO, and interestingly, it shows 2$\sigma$ deviation from the Planck cosmology \citep{Planck}.  The difference persists in the more recent Y6 analysis \citep{DES_Y6BAO2024}, which was later used to perform a joint analysis with the DES 5 Yr SNe sample \citep{DES_SNeY5_2024} and was found to support the evolving dark energy model \citep{DES_SNeBAO2025}, in agreement with the DESI result \citep{DESI_BAO1_2024,DESI_BAO2_2025}. Our analysis provides an independent means to check the DES BAO results. Besides, our result also offers a consistency check on the Y3 shear measurements, which were used to derive the Y3 shear analyses \citep{Prat_etal2022,Secco_etal2022,Amon_etal2022,y3-3x2ptkp}.

In the rest of the Letter, after presenting the theory for the GI correlation function and its covariance, we show the GI BAO measurement on a DES Y3 data sample and demonstrate its consistency with the density BAO. The calibration of the red fraction and estimation of the alignment strength and robustness tests are presented in the Appendix.

\section{ Theory}


Galaxies show intrinsic alignment (IA) \citep{TroxelIshak2015,Joachimi_alignment2015,Lamman_IAguide2024,Chisari2025}.  In the linear alignment model, galaxy shapes are stretched by the large-scale tidal field, and this IA signal can be written as \citep{Catalan_etal2001,HirataSeljak2004}
\begin{align}
  \label{eq:gamma_LA_final}
  \gamma &= - ( \partial_{xx} - \partial_{yy} , 2 \partial_{xy}  ) \nabla^{-2} \kappa ,
\end{align}
where, to align with the lensing literature, we have introduced the scalar field $\kappa $:
\beq
\kappa  \equiv  \frac{ C_1 \bar{\rho}_{\rm m 0} }{ D(z) } \delta ,
\eeq
where $C_1$ is an amplitude parameter,  $\bar{\rho}_{\rm m 0}$ is the matter density at the present time, $D$ is the linear growth factor normalized to unity at the matter-dominated era, and $\delta$ is the matter overdensity. 

We consider the cross correlation between the galaxy density field and shear field sampled by a galaxy sample.  The projected galaxy density and shear are given by
\begin{align}
  \delta_{ {\rm g} W } ( \bm{\theta} ) &=   \int d \chi W_{\rm g} (\chi) \delta_{\rm g}(\chi, \bm{\theta}), \\
  \gamma_W ( \bm{\theta} ) & =  \int d \chi W_\gamma (\chi) \gamma(\chi, \bm{\theta}) ,
\end{align}
where $  W_{\rm g}  $ and $  W_{\gamma} $ denote the tomographic windows of the galaxy density field $\delta_{\rm g}$  and the shear field $\gamma$, respectively.

Analogous to the halo-shear correlation function in galaxy-galaxy lensing \citep{dePutterTakada_2010}, the angular correlation function between the galaxy density (tomographic bin $i$, subscript g) and  radial shear field (bin $j$, subscript +), also called GI correlation function, can be written as
\begin{align}
  \label{eq:xi_gt_curvesky}
 w_{\rm g +}^{(ij)}( \theta )  = &  \sum_{l\ge2}  \frac{2l+1}{ 4 \pi l(l+1) }  \bar{P}_l^2(\cos \theta) \mathcal{C}_{\rm g \kappa}^{(ij)}(l) , 
\end{align}
where  $P_l^2$ is the associated Legendre polynomial with $m=2$, and the overbar here and hereafter signifies the average of the function over the angular bin width.  In the flat-sky limit, we have the integral approximation 
\begin{align}
  \label{eq:wgt_flatsky}
w_{\rm g +}^{(ij)}( \theta )  = &  \int \frac{dl l}{ 2 \pi }  \bar{J}_2 (l\theta) \mathcal{C}_{\rm g \kappa}^{(ij)}(l), 
\end{align}
where   $J_n$ is the cylindrical Bessel function. The angular power spectrum $  \mathcal{C}_{\rm g \kappa}^{(ij)} $ reads 
\begin{align}
  \label{eq:C_gkappa_exact}
  \mathcal{C}_{\rm g \kappa}^{ (ij)} (l)  & =  \frac{2}{\pi}  \int d k k^2   P_{\rm m }(k) b_{\rm g}^{(i)}  \int d \chi  W^{(i)}_{\rm g}(\chi) D(\chi)   j_l(k\chi)\nn \\
   & \times C_1' f_{\rm r}^{ (j) } \int d \chi'  W_{\gamma}^{(j)}  (\chi')  j_l(k\chi') , 
\end{align}
where $ C_1' \equiv C_1 \bar{\rho}_{\rm m 0} $,  $j_l$ is the spherical Bessel function,  $b_{\rm g}^{(i)}$ is the galaxy bias in the $i$th tomographic bin, and  $P_{\rm m}$ is the matter power spectrum.   Here, we have introduced the fraction of red galaxies in the sample $f_{\rm r}^{ (j) }$ to account for the fact that only the red galaxies show significant IA signals, while blue galaxies do not \citep{Hirata_etal2007,Mandelbaum_etal2011,Johnston_etal2019, Samuroff_etal2023,  HervasPeters_etal2025, Navarro-Girones_etal2025, Georgiou_etal2025, Siegel_etal2025}.  On small angular scales,  Limber approximation applies, and  Eq.~\eqref{eq:C_gkappa_exact}  reduces to 
\begin{align}  
\mathcal{C}_{\rm g \kappa}^{(ij)}(l) &=  b_{\rm g}^{(i)} f_{\rm r}^{(j)}   \int \frac{ d \chi}{\chi^2}  W_{\rm g}^{(i)} (\chi) W_\gamma^{(j)} (\chi)  C_1'  D(\chi)  P_{\rm m}\Big( \frac{ l + \frac{1}{2} }{\chi  } \Big), 
\end{align}
 We use the exact expression for $l<500$ and the Limber results for $l\ge500$.

We can derive the Gaussian covariance for the GI correlation function as in the weak lensing case (e.g., \citet{Friedrich_etal2021}). In the flat-sky limit, GI covariance reads
\begin{align}
\label{eq:Gauss_covmat}
  \mathrm{Cov} & ( w_{\rm g +}^{(ij)}( \theta ),  w^{ (mn) }_{\rm g +}( \theta' ) ) 
      =  \frac{1}{A} \int\frac{ dl l}{2 \pi} \bar{J}_2( l \theta ) \bar{J}_2( l \theta' )  \nn \\
   &  \times  \big[ \mathcal{C}_{\rm g}^{(im)}(l)  \mathcal{C}_{\kappa }^{(jn)}(l)   + \mathcal{C}_{\rm g \kappa }^{(in)}(l)  \mathcal{C}_{\rm g \kappa }^{(jm)}(l)     \big] , 
\end{align}
where $A$ is the survey area in steradians, and the power spectra  $ \mathcal{C}_{\rm g} $ and $\mathcal{C}_{\kappa} $ read 
\begin{align}
  \mathcal{C}^{(ik)}_{\rm g}(l)   =  &  \int \frac{d \chi }{ \chi^2 }  W^{(i)}_{\rm g}(\chi)  W^{(k)}_{\rm g} (\chi)  b_{\rm g}^{(i)} b_{\rm g}^{(k)} D^2  P_{\rm m } \Big(  \frac{ l + \frac{1}{2} }{\chi }  \Big)   \nn \\
   & + \delta_{ik} P_{\rm shot}   ,       \\
 \mathcal{C}^{(ik)}_{\kappa}(l)   =  &    \int \frac{d \chi }{ \chi^{ 2} } W_{\gamma}^{(i)} (\chi)   W^{(k)}_{\gamma} (\chi) f^{(i )}_{\rm r} f^{(k)}_{\rm r} C_1^{\prime 2}   P_{\rm m } \Big(  \frac{ l+ \frac{1}{2} }{\chi } \Big)  \nn \\
    & + \delta_{ik} P_{\rm shape}  , 
\end{align}
with $\delta_{ik} $ being the Kronecker delta, and $ P_{\rm shot } $  and  $ P_{\rm shape} $ denoting the shot noise of the galaxy field and the shape noise of the shear field, respectively.


We will also consider the joint BAO fit between the density and GI correlation, and so we need the cross covariance 
\begin{align}
  \label{eq:Gauss_crosscovmat}
  \mathrm{Cov}  & \big( w_{\rm g}^{(ij)} (\theta), w_{\rm g+}^{(mn)} (\theta') \big)   = 
  \frac{ 1 }{A}  \int \frac{ dl l}{ 2 \pi } \bar{J}_0(l\theta)  \bar{J}_2(l\theta')  \nn \\
 & \times  \big[  \mathcal{C}_{\rm g \kappa}^{(jn)} ( l)  \mathcal{C}_{\rm g}^{(im)}(l) 
    +    \mathcal{C}_{\rm g \kappa}^{(in)} (l) \mathcal{C}_{\rm g}^{(jm)}(l)  \big]. 
\end{align}

\section{Data, method, and results }

In the DES Y3 BAO analysis \citep{DES_Y3BAO2022,Chan_DESY3BAO2022}, out of the Y3 gold sample \citep{y3-gold}, a red galaxy sample is constructed, which is denoted as the Y3 BAO sample here \citep{y3-baosample}. Altogether, there are about 7.03 million galaxies in the photo-$z$ range $[0.6,1.1]$ \change{over a survey area of 4100 deg$^2$}. The associated random catalog samples the survey mask uniformly, with a number density 20 times that of the data catalog. Photo-$z$ is estimated by the DNF algorithm \citep{DNF}, and the characteristic photo-$z$ uncertainty is $\sim 0.03 (1+z) $. The true redshift distribution of the sample is estimated using the spectroscopic galaxies from the VIPERS survey \citep{vipers}.   For more details on the properties of the Y3 BAO sample, we refer readers to \citep{y3-baosample}, especially Table 2 therein. The associated shear properties are obtained from the full gold catalog.  The shear $\gamma$  is calibrated using the {\tt METACALIBRATION } method \citep{HuffMandelbaum2017,SheldonHuff2017}, which derives the shear from the ellipticity measurements by a response matrix. The GI correlation function is measured using {\tt TreeCorr} \citep{Jarvis_etal2004} 
\beq
\hat{w}_{\rm g + } = \langle S_+ D \rangle -  \langle S_+ R \rangle,  
\eeq
where $\langle S_+ D \rangle $ ($\langle S_+ R \rangle $) denotes the mean radial shear around the tracer galaxies (random), which reduces to the estimator in \citet{Mandelbaum_etal2006} on large scale. 

\begin{figure}[!tb]
\includegraphics[width=\linewidth]{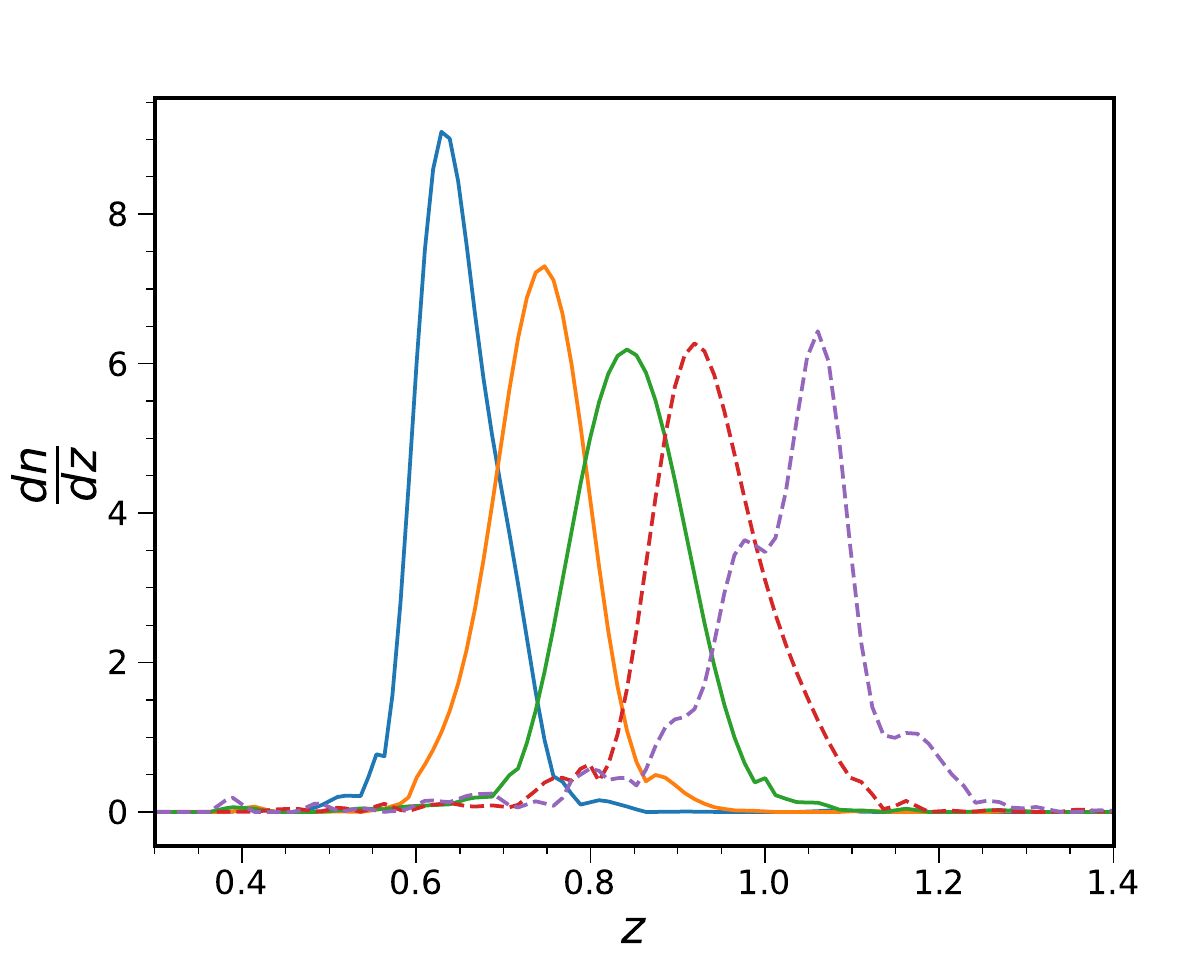}  
\caption{ True redshift distributions of the BAO sample. All five tomographic bin distributions are shown, although in the fiducial setting, only the first three bins (solid lines) are used.     }
\label{fig:Y3BAO_truez_dist}
\end{figure}

We take the same configuration as the Y3 BAO analysis,  with the data in the photo-$z$ range [0.6, 1.1] divided into five tomographic bins, each of width 0.1. However, our fiducial setting only uses the first three bins, as the last two bins barely contribute any signal.  In Fig.~\ref{fig:Y3BAO_truez_dist}, we show the true redshift distribution of the five bins.

We take the likelihood to be Gaussian,
\beq
\mathcal{L} \propto \exp \Big( -  \frac{\chi^2}{ 2 }  \Big), 
\eeq
where $\chi^2 $ is given by
\beq
\chi^2 = ( \mathcal{M} -  \mathcal{D} )^T \Sigma^{-1}  ( \mathcal{M} - \mathcal{D} ),
\eeq
with $ \mathcal{M}$ and $\mathcal{D}$, respectively, denoting the model and data vectors and $\Sigma$ being the covariance. We perform a BAO template fit using the maximum likelihood estimator following \citet{Chan_etal2018}. The full fitting model is 
\beq
\label{eq:full_model} 
\mathcal{M} = B T( \alpha \theta ) + \sum_i \frac{A_i}{\theta^i }, 
\eeq
where $T$ denotes the theory template $w_{\rm g+}$, $B$ is an amplitude parameter, and the polynomial in  $ 1/\theta $ is a broadband term. The scale dilation parameter $\alpha$  rescales the BAO scale in the transverse direction, and physically, it encodes  
\begin{align}
\alpha  & = \frac{ D_{\rm M} r_{\rm s}^{\rm fid} }{ D_{\rm M}^{\rm fid}  r_{\rm s} }, 
\end{align}
where $ D_{\rm M} $ is the comoving angular diameter distance and $r_{\rm s} $ represents the sound horizon at the drag epoch.  The quantities with (without) ``fid'' mean that they are evaluated in the fiducial (data) cosmology.


Here, we summarize the default BAO fit settings. \change{ Even though the linear alignment model is used in the fiducial setting, we incorporate BAO damping into the power spectrum \citep{Eisenstein_etal2007,CrocceScoccimarro2008,Blas_etal2016}. }  The template and the Gaussian covariance are computed in the Planck cosmology \citep{Planck}. We first take  $ C_1 \bar{\rho}_{\rm m0} = 0.04 $ \citep{ChisariDvorkin2013}, but then further adjust the amplitude with  $ f_{\rm r}$ by fitting to the $w_{\rm g+}$ measurements.  Thus, $ f_{\rm r}$ effectively accounts for both the red fraction and the alignment amplitude. See also the Appendix where, by determining the red fraction using the method in \citet{McCullough_etal2024}, we constrain the IA amplitude using the BAO strength. The data vector is measured with linear angular bin width of 0.4$^\circ$, and the fitting is performed over the angular range  [1.4, 4.6]$^\circ$.  In the default setting, we use only the first three tomographic bins and   a constant broadband term.

\begin{figure*}[!htb]
\includegraphics[width=\linewidth]{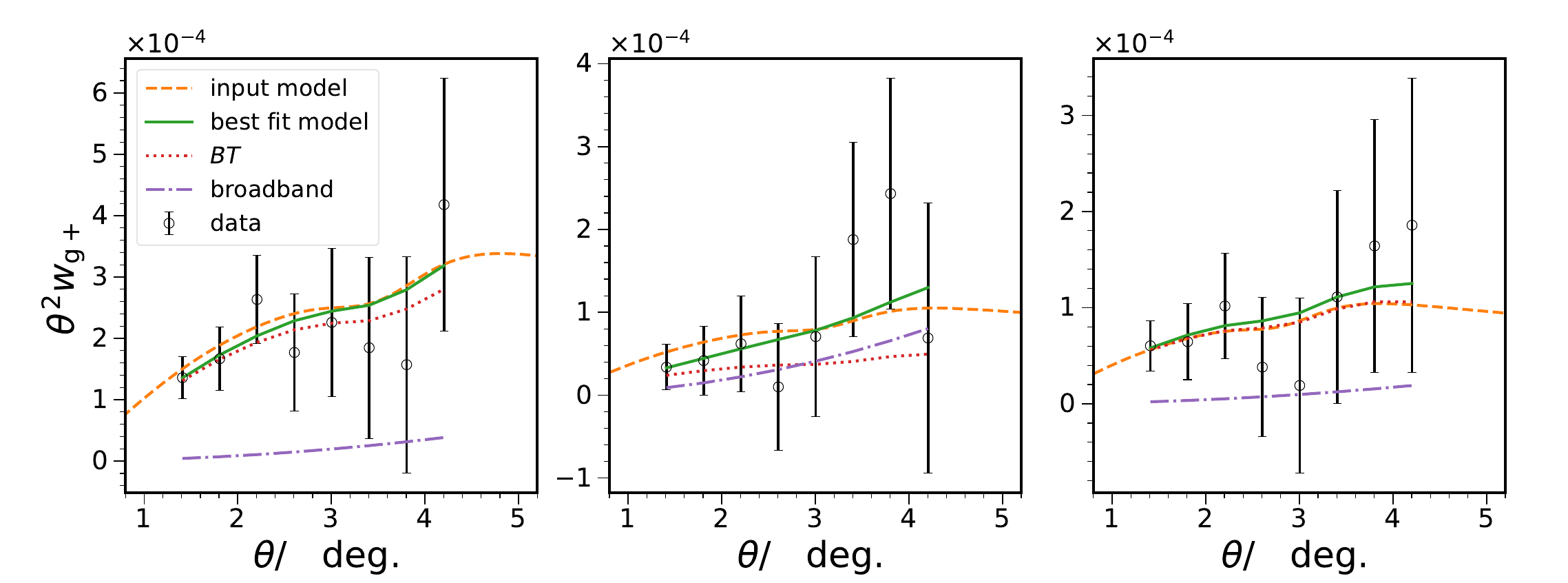}  
\caption{ Shown are the GI correlation function measurements (data points, black) and its best fit model (solid, green) for the first three tomographic bins (from left to right). We also plot the components of the best fit model, including the input template model (dashed, orange), $BT$ term (dotted, red), and broadband term (dotted-dashed, violet).   } 
\label{fig:bestfit_model_components}
\end{figure*}

In Fig.~\ref{fig:bestfit_model_components}, we show the GI correlation function measurement and its best-fit model.  The best fit  $ \alpha$ is $  0.966 \pm 0.252 $ (1$\sigma$) with a $\chi^2/{\rm dof } =  10.7/17= 0.63 $. Although the overall data are noisy, the presence of a trough at the BAO scale is apparent.  We also plot the components $BT$ and the broadband term $A_0$.  The magnitude of $B$ is close to 1 and the $A_0$ contribution is small, indicating that the fit is signal dominated. This is not the case for the last two bins (not shown here).  Small signals in those bins could be due to the poor photo-$z$ quality,  low number density \citep{y3-baosample}, and low red fraction (see the Appendix).

\begin{figure}[!htb]
\includegraphics[width=\linewidth]{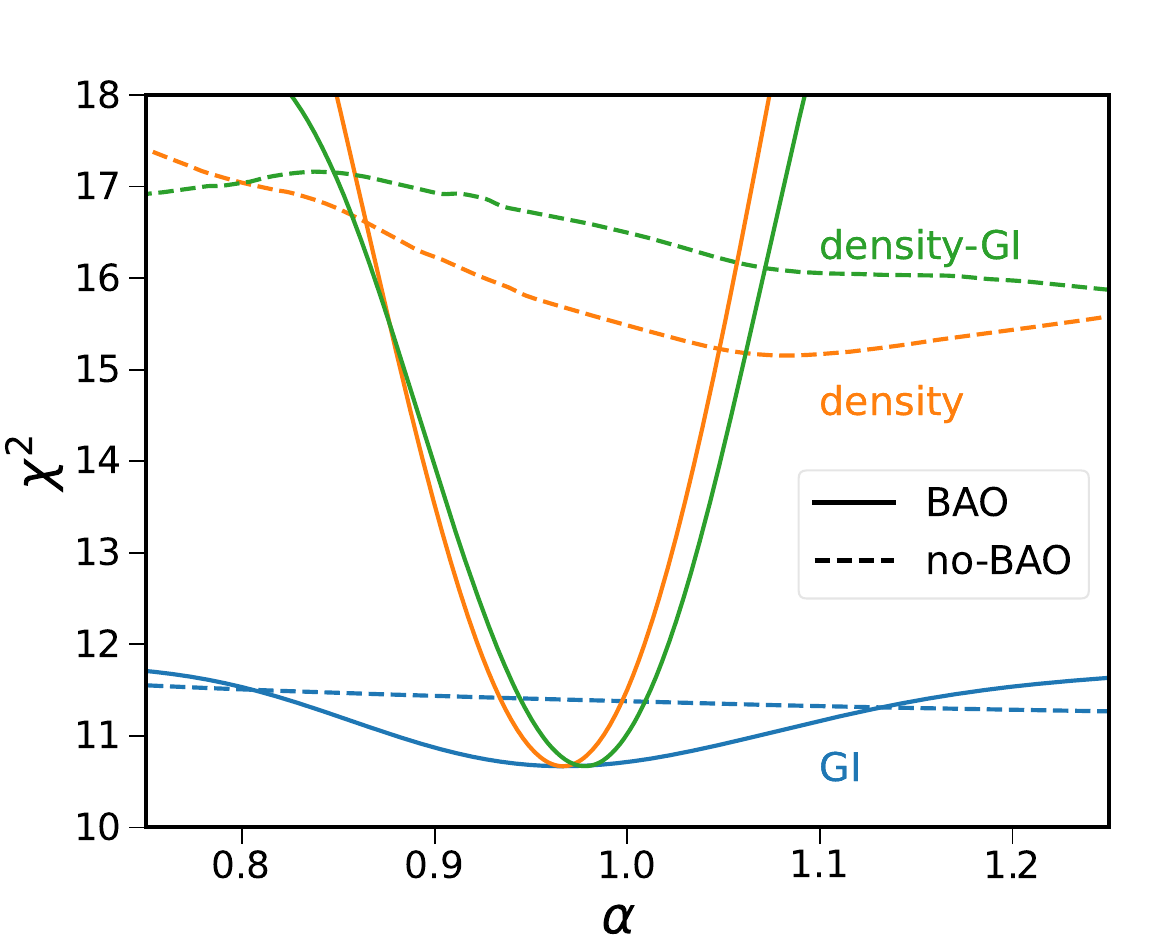}
\caption{  $\chi^2$ distribution as a function of $\alpha$, obtained with the BAO (solid) and no-BAO (dashed) model, respectively. The GI (blue), density (orange), and joint density-GI (green) BAO fit results are compared. For ease of comparison, the $\chi^2$ distributions for the density and GI-density are shifted vertically so that their BAO template results are of the same height as the GI one.
} 
\label{fig:chi2_all3_centershift}
\end{figure}

We plot the $\chi^2 $ distribution obtained with the BAO and no-BAO templates in Fig.~\ref{fig:chi2_all3_centershift}.  The profile likelihood clearly favors the BAO signature. At the best fit of the BAO model, the $\chi^2$ difference between the no-BAO model and BAO model is $\Delta \chi^2  = 0.74 $, which is equivalent to 0.86$\sigma $. In the Appendix, we test the sensitivity of the fit results to various fitting conditions and redshift bin combinations, and we find the results to be robust.

We also perform the density BAO fit following the setup in \citet{DES_Y3BAO2022}, but using only the first three tomographic bins. The resultant $\chi^2 $ distributions obtained with the BAO and no-BAO templates are also presented in Fig.~\ref{fig:chi2_all3_centershift}. The constraint on  $ \alpha$ is $  0.966 \pm 0.037 $, with a $ \chi^2 /{\rm dof} = 57.7 / 53= 1.09  $, and it is in good agreement with the GI fit result. Thus, we independently validate the DES Y3 density BAO and the shear measurements. The consistency also provides a verification for the linear alignment model on the BAO scale.


The first GI BAO detection was presented in \citet{Xu:2023vrl} using a sample derived by cross-matching the CMASS galaxies \citep{Alam_etal2017} to the DESI Legacy imaging sample \citep{Dey_etal2019}.   Despite its large volume coverage, various works (e.g.,~\citet{Phriksee_etal2020, Yao_etal2020, Zhang_etal2022}) suggest that the shear measurements from the DESI Legacy sample are not as reliable as other dedicated lensing surveys.  This may be related to the 2$\sigma$ discrepancy between the density and GI BAO fit results found in \citet{Xu:2023vrl}.

\begin{figure*}[!htb]
\includegraphics[width=\linewidth]{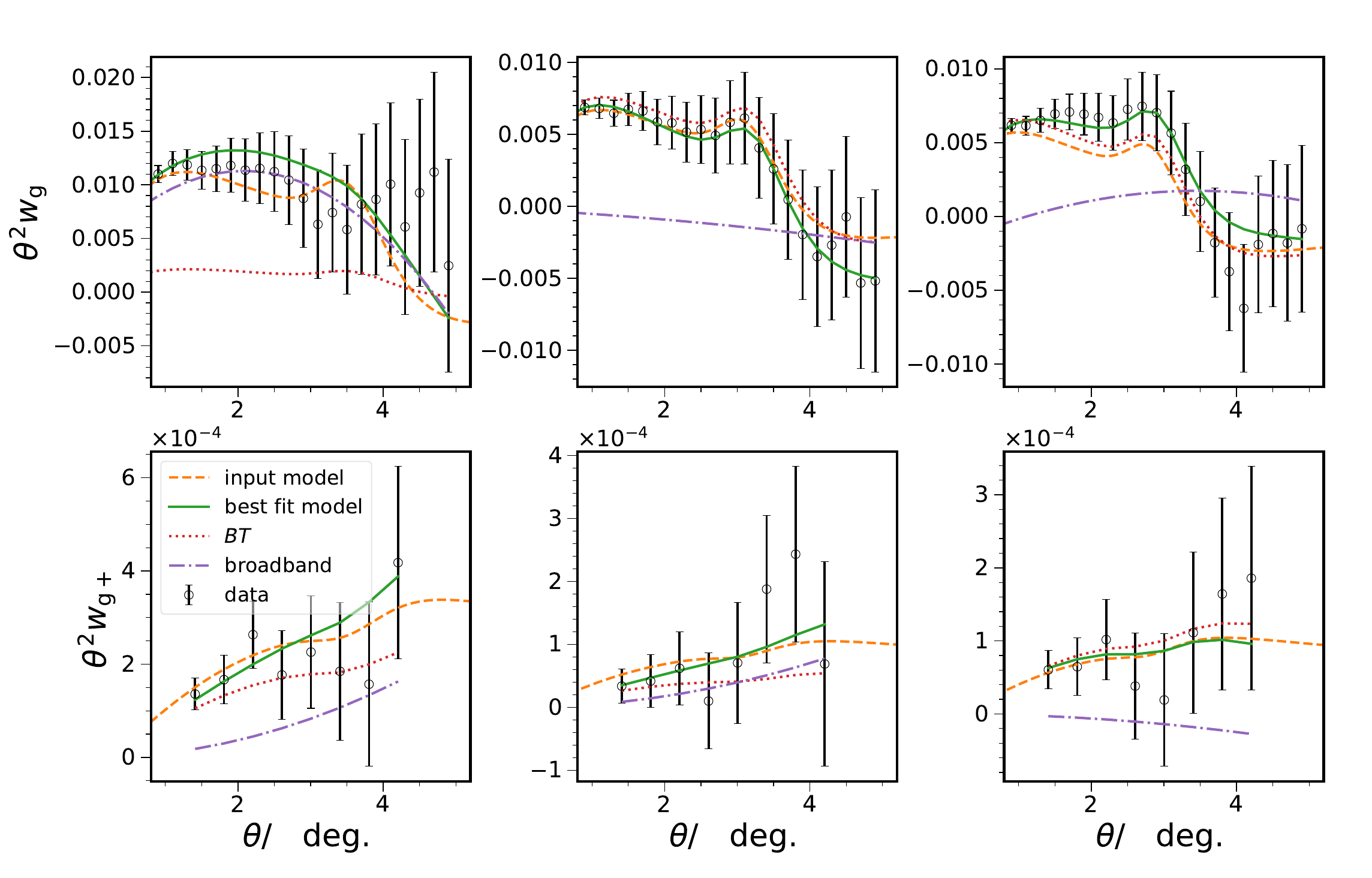}  
\caption{  The density (upper panels) and GI (lower panels) correlation functions for the first three tomographic bins (left to right) are shown.  We compare the measurements (data points, black) with the joint density-GI BAO best fit model (solid, green).  Shown also are the components of the best fit model, including the input template model (dashed, orange), $BT$ term (dotted, red), and broadband term (dotted-dashed, violet).  }
\label{fig:bestfit_densityGImodel_components}
\end{figure*}

Because our density and GI results are consistent with each other, we can safely combine them to perform a joint fit. These results are highly correlated, and so we must include their cross covariance [Eq.~\eqref{eq:Gauss_crosscovmat}].  We show the density and GI correlation function measurements in Fig.~\ref{fig:bestfit_densityGImodel_components}, together with the joint density-GI best fit model and its components.  The joint best fit yields $\alpha = 0.978 \pm 0.038 $ with a $\chi^2 /{\rm dof} = 67.5 / 71 = 0.95  $.  Interestingly, as evident in Fig.~\ref{fig:chi2_all3_centershift}, the joint best fit $\alpha$ is higher than the GI or density result alone. This is related to the bin 1 density correlation function, which exhibits a trough instead of a peak at the expected BAO position.  No systematics are found to explain it, and this pattern appears persistently in DES Y1 \citep{DES_Y1BAO2019}, Y3 \citep{DES_Y3BAO2022}, and Y6 \citep{DES_Y6BAO2024} analyses. If we fit the density BAO using bins 2 and 3 only, the best fit $ \alpha $ is $ 0.987 \pm 0.036 $.   This demonstrates that although the first bin does not show any apparent BAO signals, it does drive the best fit to a smaller $\alpha$ value.  In contrast, for the GI BAO, appreciable BAO signals are present in bin 1, but little signals are observed in bin 2.   In the joint density-GI fit, because of the presence of the BAO feature in the bin 1 GI correlation function, the influence of the first density bin is weakened.  Thus, we argue that the joint fit result is more robust to systematics effects than the individual fit results. 


\change{  In Appendix \ref{sec:S4_forecast}, we present forecasts for the GI BAO constraint in stage IV photometric surveys. The detection significance is found to range from $2\sigma$ to $5\sigma$, depending on the number density of the galaxy sample assumed. }

\section{Conclusions}

An effective way to check the consistency between datasets in cosmology is to perform cross-correlation analyses. In this work, we point out that the GI BAO is a powerful method for checking the consistency of the density BAO measurement, shear measurement, and alignment model.  The GI BAO and density BAO measurements on the same dataset should agree, as they probe the same underlying modes, with the difference that GI BAO does so as a spin-2 field.  Thus, any discrepancy would indicate untreated systematics in the measurements or modeling.  As a demonstration, we measure the GI BAO on the DES Y3 data, marking the first GI BAO measurement on a photometric data sample.  From the GI BAO, we get $ \alpha = 0.966 \pm 0.252 $, while the density BAO yields $0.966 \pm 0.037 $, in good agreement with each other. Various robustness tests for the GI BAO measurements are conducted, and we find the results to be robust.  This validates the DES Y3 BAO measurement, shear measurement, and the linear alignment model.  When combining the GI BAO with the density BAO, we find that the impact of an anomalous bin measurement in the density BAO is reduced, resulting in a more robust measurement.  GI BAO will become an even more powerful consistency check in stage IV surveys, owing to their huge data volumes.


\section{Acknowledgment}
We thank Xuanwen Guo, Jia Liu, and Teppei Okumura for useful discussions and the DES collaboration for making their data publicly available.  K.C.C. acknowledges the support by the National Science Foundation of China under grant Nos.~12273121 and 12533002, and the science research grant from the China Manned Space Project with CMS-CSST-2025-A02.  Y.L. is supported by the Major Key Project of Peng Cheng Laboratory and the National Key Research and Development Program of China under grant number 2023YFA1605600.

\appendix

\section{Calibration of the red fraction and constraint on the amplitude of IA}

Here, we explore constraining the amplitude of IA using the BAO signal. To do so, we need an independent calibration of the red fraction $f_{\rm r} $.

The Y3 BAO sample is a red galaxy sample constructed based on the wide band $riz$ information only \citep{y3-baosample}. Here, we calibrate $ f_{\rm r} $  of this sample  using the method in Appendix A of \citet{McCullough_etal2024}. With the DES deep field photometry $ugrizJHK_s $ bands, we can determine the red fraction of the galaxies in the deep field with high precision. Once we have the red fraction  in the deep field derived by the SED fitting method {\tt BAGPIPES} \citep{Carnall_etal2018}, we can propagate the results to the BAO sample by means of the transfer function computed from the {\tt BALROG} simulation \citep{Everett_etal2022}. The red fraction $f_{\rm r}$ of the five tomographic bins are determined to be [0.40, 0.39, 0.34, 0.29, 0.25].  The absolute values depend on the somewhat arbitrary definition of red galaxies adopted, but the trend strongly correlates with the alignment strength and is in agreement with several other reasonable methods used to define a red galaxy.

\begin{figure}[!tb]
  \includegraphics[width=0.5\linewidth]{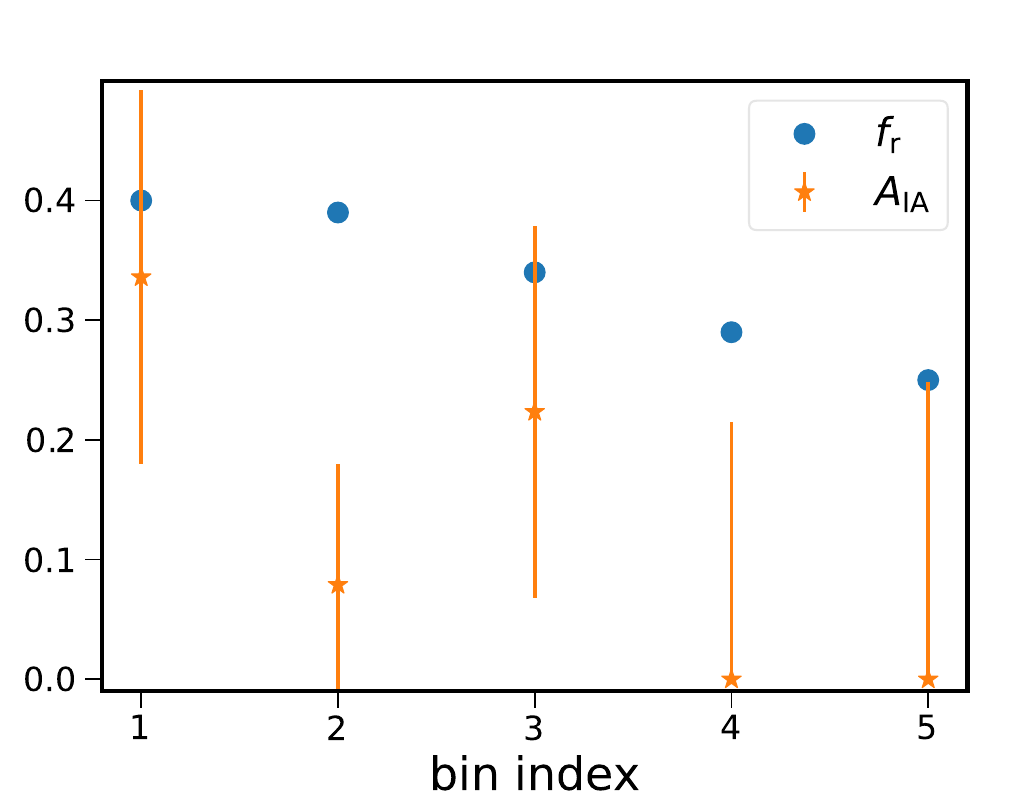}  
\caption{  The red fraction $f_{\rm r} $ (blue) and the amplitude of IA $ A_{\rm IA} $ (orange) for the five tomographic bins.  }
\label{fig:fr_AIA}
\end{figure}

In our BAO fitting pipeline, we first fit an effective parameter \( f_{\rm r} \), which is actually \( f_{\rm r} A \), where \( A \) is an amplitude parameter. Additionally, we include another amplitude parameter \( B \) in Eq.~\eqref{eq:full_model}. Consequently, the total BAO amplitude \( A_{\rm IA} \) is given by \( A B \). In Fig.~\ref{fig:fr_AIA}, we plot our estimates of \( f_{\rm r} \) and \( A_{\rm IA} \) for the five tomographic bins. We note that the trend of \( A_{\rm IA} \) in bin 2 differs from that of \( f_{\rm r} \). This discrepancy could arise if, for example, the sample in this bin contains a higher fraction of low-mass galaxies as the IA signal increases with the mass of the density tracer \citep{Fortuna_etal2025}.


\begin{table}[!tb]
	\centering
	\caption{ The upper part shows the BAO constraint under various configurations, while the lower part gives the results from different redshift bin combinations.  Two error bars for the best fit are given, corresponding to 0.5$\sigma$ and 1$\sigma$, respectively.     } 
	\label{tab:systematics_tests}
	\begin{tabular}{l  c   c }
	\hline
        Configuration                        & $ {\rm best \,  fit} \pm  0.5\sigma  \pm  1\sigma$       &    $ \chi^2/{\rm dof}  $   \\  
		\hline
     Fiducial                             &    $ 0.966  \pm 0.079 \pm 0.252 $                  &   $ 10.7/17 = 0.63 $        \\
     Y3 shear matching                    &    $ 0.930 \pm 0.072 \pm 0.197 $                   &   $ 16.8/17 = 0.99 $         \\
     
     Nonlinear alignment                  &  $ 0.957 \pm 0.078 \pm 0.255   $                   &   $ 10.7/17 =0.63   $     \\
 $ \frac{dn}{dz}$ estimation by  $z_{\rm mc}$    &   $  0.978 \pm 0.083 \pm 0.261 $                   &   $  10.7/ 17 =0.63    $   \\
     
     No broadband                            &    $ 0.968 \pm 0.074 \pm 0.204 $                   &  $ 10.9/20 =0.54 $         \\
     
     Range [1.4, 4.8]$^\circ$                 &    $ 0.950 \pm 0.073 \pm 0.185 $                   &  $ 12.7/20 = 0.64 $          \\
     Range [1.0, 4.4]$^\circ$                  &  $ 1.000 \pm 0.142  \pm --   $                    &  $ 16.1/20 = 0.81 $          \\
     MICE cosmology                          &   $ 1.025 \pm 0.071  \pm 0.173  $                  &  $ 10.5/17 = 0.62 $         \\
     Mock covariance                         &    $ 0.974 \pm 0.080 \pm 0.226 $                   &  $ 10.9/17 = 0.64 $           \\
     $ \Delta \theta = 0.2^\circ $            &   $  0.975 \pm 0.089 \pm -- $                      &  $ 48.9/38  =  1.29 $    \\         
     $ \Delta \theta = 0.6^\circ $            &   $  0.979 \pm 0.101  \pm -- $                     &  $ 9.45/8  = 1.18  $       \\         

     \hline
     $z$ bin [1]                                 &  $0.955 \pm 0.101 \pm --  $                   &  $  5.20/5=1.05 $        \\
     $z$ bin [2]                                 &  $--$                                         &        $ -- $               \\
     $z$ bin [3]                                 &  $0.947 \pm 0.139 \pm--$                      & $  2.15 / 5 = 0.43 $      \\     
     $z$ bins [1,2]                               &  $ 0.980 \pm 0.123\pm--  $                    &  $ 8.51 / 11=0.77 $       \\
     $z$ bins [2,3]                               &  $ 0.990 \pm 0.140 \pm --  $                  &  $ 5.77 /11 = 0.52 $      \\     
     $z$ bins [1,2,3,4]                           &  $ 0.966 \pm 0.079 \pm 0.252 $                &  $ 17.8/23=0.77  $     \\
     $z$ bins [1,2,3,4,5]                         &  $ 0.966 \pm 0.079 \pm 0.252 $                &  $ 26.4/29 =0.91 $     \\
     $z$ bins [3,4,5]                             &  $ 0.947 \pm 0.138 \pm -- $                   &  $ 17.9/17 = 1.05 $      \\      
     \hline
	\end{tabular}
\end{table}

\section{ Robustness tests } 
\label{Sec:systematic_tests}

We evaluate the sensitivity of the GI BAO results to different fitting conditions, which are detailed in Table \ref{tab:systematics_tests} and include the following items.
\begin{itemize}
\item The Y3 shear catalog \citep{y3-shapecatalog} used by the DES Y3 lensing analysis \citep{Prat_etal2022,Secco_etal2022,Amon_etal2022,y3-3x2ptkp} is also derived from the gold catalog \citep{y3-gold}, with a further battery of selection cuts applied to enhance the quality of the shear sample.  We find that about 90\% of the galaxies in the BAO sample can be cross-matched to this shear catalog. Because we lack independent true redshift calibration for this cross-matched sample and cross correlation is less susceptible to systematics, we adopt the full BAO sample in our fiducial analysis.  Here, we show the fit results for this shear sample, assuming the true redshift distribution from the BAO sample.  The best-fit  $\alpha =  0.930 \pm 0.197 $ is well consistent with the fiducial one.

\item \change{ In the fiducial setting, we adopt the linear alignment model with BAO damping. Here, we test the results obtained using the nonlinear alignment model \citep{BridleKing2007}. To this end, we employ the nonlinear power spectrum from {\tt camb}, computed with the halo model code {\tt HMcode-2000} \citep{Mead_etal2021}. We find that the results are insensitive to the choice of alignment model. }

\item \change{  To assess the potential impact of miscalibration in the true redshift distribution, we consider an alternative estimate of $\frac{dn}{dz}$ derived from the so-called $z_{\rm mc}$, following \citet{DES_Y3BAO2022}.  This method computes $\frac{dn }{dz } $  using the nearest spectroscopic redshift from the training sample provided by the photo-$z$ code {\tt DNF} \citep{DNF}.  Our findings indicate that the results are robust against the calibration of $\frac{dn}{dz}$. }

\item Because our constraint is relatively weak, we cannot afford to have too many degrees of freedom in the broadband term, and so in the fiducial setup we only use a constant term.  If there is no broadband term at all, the best fit is similar to the fiducial one. However, if we allow for more broadband terms, the fit will become much worse.

\item  To examine the sensitivity to the fitting range, we show the fit results in the angular range [1.4, 4.8]$^\circ$ and [1.0, 4.4]$^\circ$. When the range is extended to [1.0, 4.4]$^\circ$, the impact is relatively large, and \change{ it may suggest that for small angular scales, the degrees of freedom allowed by our limited broadband terms are not enough. }  

\item Instead of the fiducial Planck cosmology, we use the MICE cosmology \citep{MICE1} to compute the template model. We anticipate that the best-fit $\alpha$ is rescaled by a factor of 1.042 when the MICE template is used, and indeed, we find that the best-fit  $\alpha$ is well consistent with this expectation.

\item We test the theory covariance using mock catalogs, built by randomly assigning shear components to the galaxies in the DES Y3 BAO mock catalog \citep{Ferrero_etal2021}. This is possible because the shape noise is a couple of orders of magnitude larger than the shear signals. We use 960 mocks altogether. We find that the theory covariance is in good agreement with the random shear mock covariance, and the best fit $\alpha = 0.974 \pm 0.226 $ is consistent with the theory covariance result.

\item In the bin width tests, the correlation function is  measured in alternative binnings, $ \Delta \theta = 0.2^\circ  $ and $ 0.6^\circ $, and the best fits are essentially unchanged.

\item We show the fit results from various redshift bin combinations. We say that the BAO is detected if the $x \sigma$ error bounds ($x=0.5$ or 1) for $ \alpha $ fall within the interval [0.7, 1.3]; otherwise no results are shown. Single bin fit results  with bin 1 or 3 only are consistent with the fiducial fit, but the second bin result is much weaker, and the 0.5$\sigma$ error bar cannot be defined. Other bin combinations also show consistent results.  Bin 4 or 5 do not show the BAO signal because their best-fit $B$ values in both cases are zero. Thus, in the fiducial setting, we only use the first three bins. 
  

\end{itemize}

\section{ Forecast for Stage IV Surveys  } 
\label{sec:S4_forecast}

A large amount of photometric data will be available from stage IV surveys such as Rubin \citep{LSST_2009} and CSST \citep{CSST_2026}. Here, we forecast the significance of the GI BAO detection in these upcoming surveys.

The forecast is set up as follows. The survey area is 18,000 $\mathrm{deg}^2 $. The true underlying distribution of the galaxy sample is assumed to be \citep{LSST_DE_2018}
\beq
\frac{dn }{ dz} \propto z^2 \exp \Big[ -\Big( \frac{ z }{z_0 } \Big)^\alpha    \big], 
\eeq
with $z_0=0.28$ and $\alpha = 0.90 $.  The photo-$z$ uncertainty is taken to be $0.03(1+z)$,  and the true-$z$ distribution is assumed to be Gaussian. We consider six tomographic bins of equal width  $\Delta z = 0.1 $ over the redshift range $[0.4, 1]$.  The shape noise is assumed to be at the same level as DES Y3,  $\sigma_\kappa=0.2$.  The bias parameter is set to 1.8. We consider two galaxy samples: a sparse sample amounting to 3 galaxies per arcmin$^2$  and a dense sample amounting to 11.6 galaxies per arcmin$^2$ in the redshift range [0.4,1].  For comparison, the DES Y6 BAO sample contains 1.0 galaxy per arcmin$^2$.

Employing the parameters described above, we generate the model, data vector, and covariance matrix. The resulting best-fit error bars are then used to forecast the GI BAO constraints.  For the sparse sample, the $1\sigma$ GI BAO constraint yields $\alpha = 1.000 \pm 0.058$, corresponding to a detection significance of $2.0\sigma$. For the dense sample, the constraint tightens to $\alpha = 0.999 \pm 0.022$, with an associated detection significance of $4.7\sigma$.

\bibliographystyle{aasjournal}
\bibliography{references}

@ARTICLE{LSST_DE_2018,
       author = {{The LSST Dark Energy Science Collaboration} and {Mandelbaum}, Rachel and {Eifler}, Tim and {Hlo{\v{z}}ek}, Ren{\'e}e and {Collett}, Thomas and {Gawiser}, Eric and {Scolnic}, Daniel and {Alonso}, David and {Awan}, Humna and {Biswas}, Rahul and {Blazek}, Jonathan and {Burchat}, Patricia and {Chisari}, Nora Elisa and {Dell'Antonio}, Ian and {Digel}, Seth and {Frieman}, Josh and {Goldstein}, Daniel A. and {Hook}, Isobel and {Ivezi{\'c}}, {\v{Z}}eljko and {Kahn}, Steven M. and {Kamath}, Sowmya and {Kirkby}, David and {Kitching}, Thomas and {Krause}, Elisabeth and {Leget}, Pierre-Fran{\c{c}}ois and {Marshall}, Philip J. and {Meyers}, Joshua and {Miyatake}, Hironao and {Newman}, Jeffrey A. and {Nichol}, Robert and {Rykoff}, Eli and {Sanchez}, F. Javier and {Slosar}, An{\v{z}}e and {Sullivan}, Mark and {Troxel}, M.~A.},
        title = "{The LSST Dark Energy Science Collaboration (DESC) Science Requirements Document}",
      journal = {arXiv e-prints},
         year = 2018,
        month = sep,
          eid = {arXiv:1809.01669},
        pages = {arXiv:1809.01669},
          doi = {10.48550/arXiv.1809.01669},
archivePrefix = {arXiv},
       eprint = {1809.01669},
 primaryClass = {astro-ph.CO},
       adsurl = {https://ui.adsabs.harvard.edu/abs/2018arXiv180901669T}
}

@ARTICLE{CSST_2026,
       author = {{CSST Collaboration} and {Gong}, Yan and {Miao}, Haitao and {Zhan}, Hu and {Li}, Zhao-Yu and {Shangguan}, Jinyi and {Li}, Haining and {Liu}, Chao and {Chen}, Xuefei and {Yuan}, Haibo and {Zhou}, Jilin and {Liu}, Hui-Gen and {Yu}, Cong and {Ji}, Jianghui and {Qi}, Zhaoxiang and {Liu}, Jiacheng and {Dai}, Zigao and {Wang}, Xiaofeng and {Zheng}, Zhenya and {Hao}, Lei and {Dou}, Jiangpei and {Ao}, Yiping and {Lin}, Zhenhui and {Zhang}, Kun and {Wang}, Wei and {Sun}, Guotong and {Li}, Ran and {Li}, Guoliang and {Xu}, Youhua and {Li}, Xinfeng and {Li}, Shengyang and {Wu}, Peng and {Zhang}, Jiuxing and {Wang}, Bo and {Bai}, Jinming and {Cai}, Yi-Fu and {Cai}, Zheng and {Cao}, Jie and {Chan}, Kwan Chuen and {Chang}, Jin and {Chen}, Xiaodian and {Chen}, Xuelei and {Chen}, Yuqin and {Chen}, Yun and {Cui}, Wei and {Dong}, Subo and {Du}, Pu and {Duan}, Wenying and {Fan}, Junhui and {Fan}, LuLu and {Fan}, Zhou and {Fan}, Zuhui and {Fang}, Taotao and {Fu}, Jianning and {Fu}, Liping and {Fu}, Zhensen and {Gao}, Jian and {Gu}, Shenghong and {Gu}, Yidong and {Guo}, Qi and {Han}, Zhanwen and {Hu}, Bin and {Huang}, Zhiqi and {Ho}, Luis C. and {Jiang}, Linhua and {Jiang}, Ning and {Jing}, Yipeng and {Kang}, Xi and {Kong}, Xu and {Li}, Cheng and {Li}, Chengyuan and {Li}, Di and {Li}, Jing and {Li}, Nan and {Li}, Yang A. and {Liao}, Shilong and {Lin}, Weipeng and {Liu}, Fengshan and {Liu}, Jifeng and {Liu}, Xiangkun and {Liu}, Zhuokai and {Mao}, Ruiqing and {Mao}, Shude and {Meng}, Xianmin and {Pang}, Xiaoying and {Peng}, Xiyan and {Peng}, Yingjie and {Shan}, Huanyuan and {Shen}, Juntai and {Shen}, Shiyin and {Shen}, Zhiqiang and {Shi}, Sheng-Cai and {Shi}, Yong and {Tan}, Siyuan and {Tian}, Hao and {Wang}, Jianmin and {Wang}, Jun-Xian and {Wang}, Xin and {Wang}, Yuting and {Wu}, Hong and {Wu}, Jingwen and {Wu}, Xuebing and {Xu}, Chun and {Xue}, Xiang-Xiang and {Xue}, Yongquan and {Yang}, Ji and {Yang}, Xiaohu and {Yao}, Qijun and {Yuan}, Fangting and {Yuan}, Zhen and {Zhang}, Jun and {Zhang}, Pengjie and {Zhang}, Tianmeng and {Zhang}, Wei and {Zhang}, Xin and {Zhao}, Gang and {Zhao}, Gongbo and {Zhong}, Hongen and {Zhong}, Jing and {Zhou}, Liyong and {Zhu}, Wei and {Zu}, Ying},
        title = "{Introduction to the Chinese Space Station Survey Telescope (CSST)}",
      journal = {Science China Physics, Mechanics, and Astronomy},
         year = 2026,
        month = jan,
       volume = {69},
       number = {3},
          eid = {239501},
        pages = {239501},
          doi = {10.1007/s11433-025-2809-0},
archivePrefix = {arXiv},
       eprint = {2507.04618},
 primaryClass = {astro-ph.IM},
       adsurl = {https://ui.adsabs.harvard.edu/abs/2026SCPMA..6939501C}
}

@article{LSST_2009,
	adsurl = {https://ui.adsabs.harvard.edu/abs/2009arXiv0912.0201L},
	archiveprefix = {arXiv},
	author = {{LSST Science Collaboration}},
	eid = {arXiv:0912.0201},
	eprint = {0912.0201},
	journal = {arXiv e-prints},
	month = dec,
	pages = {arXiv:0912.0201},
	primaryclass = {astro-ph.IM},
	title = {{LSST Science Book, Version 2.0}},
	year = 2009}

@ARTICLE{Mead_etal2021,
       author = {{Mead}, A.~J. and {Brieden}, S. and {Tr{\"o}ster}, T. and {Heymans}, C.},
        title = "{HMCODE-2020: improved modelling of non-linear cosmological power spectra with baryonic feedback}",
      journal = {\mnras},
         year = 2021,
        month = mar,
       volume = {502},
       number = {1},
        pages = {1401-1422},
          doi = {10.1093/mnras/stab082},
archivePrefix = {arXiv},
       eprint = {2009.01858},
 primaryClass = {astro-ph.CO},
       adsurl = {https://ui.adsabs.harvard.edu/abs/2021MNRAS.502.1401M}
}

@ARTICLE{Blas_etal2016,
       author = {{Blas}, Diego and {Garny}, Mathias and {Ivanov}, Mikhail M. and {Sibiryakov}, Sergey},
        title = "{Time-sliced perturbation theory II: baryon acoustic oscillations and infrared resummation}",
      journal = {\jcap},
         year = 2016,
        month = jul,
       volume = {2016},
       number = {7},
          eid = {028},
        pages = {028},
          doi = {10.1088/1475-7516/2016/07/028},
archivePrefix = {arXiv},
       eprint = {1605.02149},
 primaryClass = {astro-ph.CO},
       adsurl = {https://ui.adsabs.harvard.edu/abs/2016JCAP...07..028B}
}

@ARTICLE{Eisenstein_etal2007,
       author = {{Eisenstein}, Daniel J. and {Seo}, Hee-Jong and {White}, Martin},
        title = "{On the Robustness of the Acoustic Scale in the Low-Redshift Clustering of Matter}",
      journal = {\apj},
         year = 2007,
        month = aug,
       volume = {664},
       number = {2},
        pages = {660-674},
          doi = {10.1086/518755},
archivePrefix = {arXiv},
       eprint = {astro-ph/0604361},
 primaryClass = {astro-ph},
       adsurl = {https://ui.adsabs.harvard.edu/abs/2007ApJ...664..660E}
}

@ARTICLE{CrocceScoccimarro2008,
       author = {{Crocce}, Mart{\'\i}n and {Scoccimarro}, Rom{\'a}n},
        title = "{Nonlinear evolution of baryon acoustic oscillations}",
      journal = {\prd},
         year = 2008,
        month = jan,
       volume = {77},
       number = {2},
          eid = {023533},
        pages = {023533},
          doi = {10.1103/PhysRevD.77.023533},
archivePrefix = {arXiv},
       eprint = {0704.2783},
 primaryClass = {astro-ph},
       adsurl = {https://ui.adsabs.harvard.edu/abs/2008PhRvD..77b3533C}
}

@ARTICLE{Fortuna_etal2025,
       author = {{Fortuna}, Maria Cristina and {Dvornik}, Andrej and {Hoekstra}, Henk and {Chisari}, Nora Elisa and {Asgari}, Marika and others },
        title = "{KiDS-1000: Weak lensing and intrinsic alignment around luminous red galaxies}",
      journal = {\aap},
         year = 2025,
        month = feb,
       volume = {694},
          eid = {A322},
        pages = {A322},
          doi = {10.1051/0004-6361/202452347},
archivePrefix = {arXiv},
       eprint = {2409.15416},
 primaryClass = {astro-ph.CO},
       adsurl = {https://ui.adsabs.harvard.edu/abs/2025A&A...694A.322F}
}

@ARTICLE{Porredon_etal2025,
       author = {{Porredon}, A. and {Blake}, C. and {Lange}, J.~U. and {Emas}, N. and {Aguilar}, J. and others},
        title = "{DESI-DR1 $3 \times 2$-pt analysis: consistent cosmology across weak lensing surveys}",
      journal = {arXiv e-prints},
         year = 2025,
        month = dec,
          eid = {arXiv:2512.15960},
        pages = {arXiv:2512.15960},
          doi = {10.48550/arXiv.2512.15960},
archivePrefix = {arXiv},
       eprint = {2512.15960},
 primaryClass = {astro-ph.CO},
       adsurl = {https://ui.adsabs.harvard.edu/abs/2025arXiv251215960P}
}

@ARTICLE{ Piccirilli_etal2025,
       author = {{Piccirilli}, Giulia and {Zennaro}, Matteo and {Garc{\'\i}a-Garc{\'\i}a}, Carlos and {Alonso}, David},
        title = "{Robust cosmic shear with small-scale nulling}",
      journal = {\jcap},
         year = 2025,
        month = oct,
       volume = {2025},
       number = {10},
          eid = {017},
        pages = {017},
          doi = {10.1088/1475-7516/2025/10/017},
archivePrefix = {arXiv},
       eprint = {2502.17339},
 primaryClass = {astro-ph.CO},
       adsurl = {https://ui.adsabs.harvard.edu/abs/2025JCAP...10..017P}
}

@ARTICLE{Bigwood_etal2024,
       author = {{Bigwood}, L. and {Amon}, A. and {Schneider}, A. and {Salcido}, J. and {McCarthy}, I.~G. and others },
        title = "{Weak lensing combined with the kinetic Sunyaev-Zel'dovich effect: a study of baryonic feedback}",
      journal = {\mnras},
         year = 2024,
        month = oct,
       volume = {534},
       number = {1},
        pages = {655-682},
          doi = {10.1093/mnras/stae2100},
archivePrefix = {arXiv},
       eprint = {2404.06098},
 primaryClass = {astro-ph.CO},
       adsurl = {https://ui.adsabs.harvard.edu/abs/2024MNRAS.534..655B}
}

@ARTICLE{Siegel_etal2025,
       author = {{Siegel}, J. and {McCullough}, J. and {Amon}, A. and {Lamman}, C. and {Jeffrey}, N. and others },
        title = "{Intrinsic alignment demographics for next-generation lensing: Revealing galaxy property trends with DESI Y1 direct measurements}",
      journal = {arXiv e-prints},
         year = 2025,
        month = jul,
          eid = {arXiv:2507.11530},
        pages = {arXiv:2507.11530},
          doi = {10.48550/arXiv.2507.11530},
archivePrefix = {arXiv},
       eprint = {2507.11530},
 primaryClass = {astro-ph.CO},
       adsurl = {https://ui.adsabs.harvard.edu/abs/2025arXiv250711530S}
}

@ARTICLE{ Navarro-Girones_etal2025,
       author = {{Navarro-Giron{\'e}s}, D. and {Crocce}, M. and {Gazta{\~n}aga}, E. and {Wittje}, A. and {Siudek}, M. and others },
        title = "{The PAU Survey: Measuring intrinsic galaxy alignments in deep wide fields as a function of colour, luminosity, stellar mass and redshift}",
      journal = {arXiv e-prints},
         year = 2025,
        month = may,
          eid = {arXiv:2505.15470},
        pages = {arXiv:2505.15470},
          doi = {10.48550/arXiv.2505.15470},
archivePrefix = {arXiv},
       eprint = {2505.15470},
 primaryClass = {astro-ph.CO},
       adsurl = {https://ui.adsabs.harvard.edu/abs/2025arXiv250515470N}
}

@ARTICLE{HervasPeters_etal2025,
       author = {{Hervas Peters}, Fabian and {Kilbinger}, Martin and {Paviot}, Romain and {Baumont}, Lucie and {Russier}, Elisa and others },
        title = "{UNIONS: A direct measurement of intrinsic alignment with BOSS/eBOSS spectroscopy}",
      journal = {\aap},
         year = 2025,
        month = jul,
       volume = {699},
          eid = {A201},
        pages = {A201},
          doi = {10.1051/0004-6361/202453442},
archivePrefix = {arXiv},
       eprint = {2412.01790},
 primaryClass = {astro-ph.CO},
       adsurl = {https://ui.adsabs.harvard.edu/abs/2025A&A...699A.201H}
}

@ARTICLE{Samuroff_etal2023,
       author = {{Samuroff}, S. and {Mandelbaum}, R. and {Blazek}, J. and {Campos}, A. and {MacCrann}, N. and others },
        title = "{The Dark Energy Survey Year 3 and eBOSS: constraining galaxy intrinsic alignments across luminosity and colour space}",
      journal = {\mnras},
         year = 2023,
        month = sep,
       volume = {524},
       number = {2},
        pages = {2195-2223},
          doi = {10.1093/mnras/stad2013},
archivePrefix = {arXiv},
       eprint = {2212.11319},
 primaryClass = {astro-ph.CO},
       adsurl = {https://ui.adsabs.harvard.edu/abs/2023MNRAS.524.2195S}
}

@ARTICLE{ Georgiou_etal2025,
       author = {{Georgiou}, Christos and {Chisari}, Nora Elisa and {Bilicki}, Maciej and {La Barbera}, Francesco and {Napolitano}, Nicola R. and others },
        title = "{Intrinsic galaxy alignments in the KiDS-1000 bright sample: Dependence on colour, luminosity, morphology, and galaxy scale}",
      journal = {\aap},
         year = 2025,
        month = jul,
       volume = {699},
          eid = {A252},
        pages = {A252},
          doi = {10.1051/0004-6361/202554134},
archivePrefix = {arXiv},
       eprint = {2502.09452},
 primaryClass = {astro-ph.CO},
       adsurl = {https://ui.adsabs.harvard.edu/abs/2025A&A...699A.252G}
}

@ARTICLE{Carnall_etal2018,
       author = {{Carnall}, A.~C. and {McLure}, R.~J. and {Dunlop}, J.~S. and {Dav{\'e}}, R.},
        title = "{Inferring the star formation histories of massive quiescent galaxies with BAGPIPES: evidence for multiple quenching mechanisms}",
      journal = {\mnras},
         year = 2018,
        month = nov,
       volume = {480},
       number = {4},
        pages = {4379-4401},
          doi = {10.1093/mnras/sty2169},
archivePrefix = {arXiv},
       eprint = {1712.04452},
 primaryClass = {astro-ph.GA},
       adsurl = {https://ui.adsabs.harvard.edu/abs/2018MNRAS.480.4379C}
}

@ARTICLE{Everett_etal2022,
       author = {{Everett}, S. and {Yanny}, B. and {Kuropatkin}, N. and {Huff}, E.~M. and {Zhang}, Y. and others},
        title = "{Dark Energy Survey Year 3 Results: Measuring the Survey Transfer Function with Balrog}",
      journal = {\apjs},
         year = 2022,
        month = jan,
       volume = {258},
       number = {1},
          eid = {15},
        pages = {15},
          doi = {10.3847/1538-4365/ac26c1},
archivePrefix = {arXiv},
       eprint = {2012.12825},
 primaryClass = {astro-ph.CO},
       adsurl = {https://ui.adsabs.harvard.edu/abs/2022ApJS..258...15E}
}

@ARTICLE{McCullough_etal2024,
       author = {{McCullough}, J. and {Amon}, A. and {Legnani}, E. and {Gruen}, D. and {Roodman}, A. and others},
        title = "{Dark Energy Survey Year 3: Blue Shear}",
      journal = {arXiv e-prints},
         year = 2024,
        month = oct,
          eid = {arXiv:2410.22272},
        pages = {arXiv:2410.22272},
          doi = {10.48550/arXiv.2410.22272},
archivePrefix = {arXiv},
       eprint = {2410.22272},
 primaryClass = {astro-ph.CO},
       adsurl = {https://ui.adsabs.harvard.edu/abs/2024arXiv241022272M}
}

@ARTICLE{Hirata_etal2007,
       author = {{Hirata}, Christopher M. and {Mandelbaum}, Rachel and {Ishak}, Mustapha and {Seljak}, Uro{\v{s}} and {Nichol}, Robert and others },
        title = "{Intrinsic galaxy alignments from the 2SLAQ and SDSS surveys: luminosity and redshift scalings and implications for weak lensing surveys}",
      journal = {\mnras},
         year = 2007,
        month = nov,
       volume = {381},
       number = {3},
        pages = {1197-1218},
          doi = {10.1111/j.1365-2966.2007.12312.x},
archivePrefix = {arXiv},
       eprint = {astro-ph/0701671},
 primaryClass = {astro-ph},
       adsurl = {https://ui.adsabs.harvard.edu/abs/2007MNRAS.381.1197H}
}

@ARTICLE{Mandelbaum_etal2011,
       author = {{Mandelbaum}, Rachel and {Blake}, Chris and {Bridle}, Sarah and {Abdalla}, Filipe B. and {Brough}, Sarah and others },
        title = "{The WiggleZ Dark Energy Survey: direct constraints on blue galaxy intrinsic alignments at intermediate redshifts}",
      journal = {\mnras},
         year = 2011,
        month = jan,
       volume = {410},
       number = {2},
        pages = {844-859},
          doi = {10.1111/j.1365-2966.2010.17485.x},
archivePrefix = {arXiv},
       eprint = {0911.5347},
 primaryClass = {astro-ph.CO},
       adsurl = {https://ui.adsabs.harvard.edu/abs/2011MNRAS.410..844M}
}

@ARTICLE{Johnston_etal2019,
       author = {{Johnston}, Harry and {Georgiou}, Christos and {Joachimi}, Benjamin and {Hoekstra}, Henk and {Chisari}, Nora Elisa and others },
        title = "{KiDS+GAMA: Intrinsic alignment model constraints for current and future weak lensing cosmology}",
      journal = {\aap},
         year = 2019,
        month = apr,
       volume = {624},
          eid = {A30},
        pages = {A30},
          doi = {10.1051/0004-6361/201834714},
archivePrefix = {arXiv},
       eprint = {1811.09598},
 primaryClass = {astro-ph.CO},
       adsurl = {https://ui.adsabs.harvard.edu/abs/2019A&A...624A..30J}
}

@ARTICLE{BridleKing2007,
       author = {{Bridle}, Sarah and {King}, Lindsay},
        title = "{Dark energy constraints from cosmic shear power spectra: impact of intrinsic alignments on photometric redshift requirements}",
      journal = {New Journal of Physics},
         year = 2007,
        month = dec,
       volume = {9},
       number = {12},
        pages = {444},
          doi = {10.1088/1367-2630/9/12/444},
archivePrefix = {arXiv},
       eprint = {0705.0166},
 primaryClass = {astro-ph},
       adsurl = {https://ui.adsabs.harvard.edu/abs/2007NJPh....9..444B}
}

@article{Planck,
	adsurl = {https://ui.adsabs.harvard.edu/abs/2020A&A...641A...6P},
	archiveprefix = {arXiv},
	author = {{Planck Collaboration}},
	doi = {10.1051/0004-6361/201833910},
	eid = {A6},
	eprint = {1807.06209},
	journal = {\aap},
	month = sep,
	pages = {A6},
	primaryclass = {astro-ph.CO},
	title = {{Planck 2018 results. VI. Cosmological parameters}},
	volume = {641},
	year = 2020}

@ARTICLE{PerivolaropoulosSkara_2022,
       author = {{Perivolaropoulos}, L. and {Skara}, F.},
        title = "{Challenges for LambdaCDM: An update}",
      journal = { New Astron.~Rev. },
         year = 2022,
        month = dec,
       volume = {95},
          eid = {101659},
        pages = {101659},
          doi = {10.1016/j.newar.2022.101659},
archivePrefix = {arXiv},
       eprint = {2105.05208},
 primaryClass = {astro-ph.CO},
       adsurl = {https://ui.adsabs.harvard.edu/abs/2022NewAR..9501659P}
}

@ARTICLE{DiValentino_etal2025,
       author = {{Di Valentino}, Eleonora and {Said}, Jackson Levi and {Riess}, Adam and {Pollo}, Agnieszka and {Poulin}, Vivian and  others},
        title = "{The CosmoVerse White Paper: Addressing observational tensions in cosmology with systematics and fundamental physics}",
      journal = {Physics of the Dark Universe},
         year = 2025,
        month = sep,
       volume = {49},
          eid = {101965},
        pages = {101965},
          doi = {10.1016/j.dark.2025.101965},
archivePrefix = {arXiv},
       eprint = {2504.01669},
 primaryClass = {astro-ph.CO},
       adsurl = {https://ui.adsabs.harvard.edu/abs/2025PDU....4901965D}
}

@ARTICLE{Verde_etal2024,
       author = {{Verde}, Licia and {Sch{\"o}neberg}, Nils and {Gil-Mar{\'\i}n}, H{\'e}ctor},
        title = "{A Tale of Many H $_{0}$}",
      journal = {\araa},
         year = 2024,
        month = sep,
       volume = {62},
       number = {1},
        pages = {287-331},
          doi = {10.1146/annurev-astro-052622-033813},
archivePrefix = {arXiv},
       eprint = {2311.13305},
 primaryClass = {astro-ph.CO},
       adsurl = {https://ui.adsabs.harvard.edu/abs/2024ARA&A..62..287V}
}

@ARTICLE{DES_Y63x2pt2026,
       author = { { DES Collaboration} },
        title = "{Dark Energy Survey Year 6 Results: Cosmological Constraints from Galaxy Clustering and Weak Lensing}",
      journal = {arXiv e-prints},
         year = 2026,
        month = jan,
          eid = {arXiv:2601.14559},
        pages = {arXiv:2601.14559},
          doi = {10.48550/arXiv.2601.14559},
archivePrefix = {arXiv},
       eprint = {2601.14559},
 primaryClass = {astro-ph.CO},
       adsurl = {https://ui.adsabs.harvard.edu/abs/2026arXiv260114559D}
}

@ARTICLE{Wright_etal2025,
       author = {{Wright}, Angus H. and {St{\"o}lzner}, Benjamin and {Asgari}, Marika and {Bilicki}, Maciej and {Giblin}, Benjamin and others },
        title = "{KiDS-Legacy: Cosmological constraints from cosmic shear with the complete Kilo-Degree Survey}",
      journal = {\aap},
         year = 2025,
        month = nov,
       volume = {703},
          eid = {A158},
        pages = {A158},
          doi = {10.1051/0004-6361/202554908},
archivePrefix = {arXiv},
       eprint = {2503.19441},
 primaryClass = {astro-ph.CO},
       adsurl = {https://ui.adsabs.harvard.edu/abs/2025A&A...703A.158W}
}

@ARTICLE{SunyaevZeldovich1970,
   author = {{Sunyaev}, R.~A. and {Zeldovich}, Y.~B.},
    title = "{Small-Scale Fluctuations of Relic Radiation}",
  journal = {Astrophysics and Space Science},
     year = 1970,
    month = apr,
   volume = 7,
    pages = {3-19},
      doi = {10.1007/BF00653471},
   adsurl = {http://adsabs.harvard.edu/abs/1970Ap%26SS...7....3S}
}

@ARTICLE{PeeblesYu1970,
   author = {{Peebles}, P.~J.~E. and {Yu}, J.~T.},
    title = "{Primeval Adiabatic Perturbation in an Expanding Universe}",
  journal = {ApJ},
     year = 1970,
    month = dec,
   volume = 162,
    pages = {815},
      doi = {10.1086/150713},
   adsurl = {http://adsabs.harvard.edu/abs/1970ApJ...162..815P}
}

@ARTICLE{Weinberg_etal2013,
       author = {{Weinberg}, David H. and {Mortonson}, Michael J. and {Eisenstein}, Daniel J. and {Hirata}, Christopher and {Riess}, Adam G. and {Rozo}, Eduardo},
        title = "{Observational probes of cosmic acceleration}",
      journal = {\physrep},
         year = 2013,
        month = sep,
       volume = {530},
       number = {2},
        pages = {87-255},
          doi = {10.1016/j.physrep.2013.05.001},
archivePrefix = {arXiv},
       eprint = {1201.2434},
 primaryClass = {astro-ph.CO},
       adsurl = {https://ui.adsabs.harvard.edu/abs/2013PhR...530...87W}
}

@ARTICLE{Eisenstein_etal2005,
   author = {{Eisenstein}, D.~J. and {Zehavi}, I. and {Hogg}, D.~W. and {Scoccimarro}, R. and 
	{Blanton}, M.~R. and others },
    title = "{Detection of the Baryon Acoustic Peak in the Large-Scale Correlation Function of SDSS Luminous Red Galaxies}",
  journal = {ApJ},
   eprint = {astro-ph/0501171},
     year = 2005,
    month = nov,
   volume = 633,
    pages = {560-574},
      doi = {10.1086/466512},
   adsurl = {http://adsabs.harvard.edu/abs/2005ApJ...633..560E}
}

@ARTICLE{Cole_etal2005,
   author = {{Cole}, S. and {Percival}, W.~J. and {Peacock}, J.~A. and {Norberg}, P. and 
	{Baugh}, C.~M. and others },
    title = "{The 2dF Galaxy Redshift Survey: power-spectrum analysis of the final data set and cosmological implications}",
  journal = {MNRAS},
   eprint = {astro-ph/0501174},
     year = 2005,
    month = sep,
   volume = 362,
    pages = {505-534},
      doi = {10.1111/j.1365-2966.2005.09318.x},
   adsurl = {http://adsabs.harvard.edu/abs/2005MNRAS.362..505C}
}

@article{Gaztanaga:2008xz,
      author         = "Gaztanaga, Enrique and Cabre, Anna and Hui, Lam",
      title          = "{Clustering of Luminous Red Galaxies IV: Baryon Acoustic
                        Peak in the Line-of-Sight Direction and a Direct
                        Measurement of H(z)}",
      journal        = "MNRAS",
      volume         = "399",
      year           = "2009",
      pages          = "1663-1680",
      doi            = "10.1111/j.1365-2966.2009.15405.x",
      eprint         = "0807.3551",
      archivePrefix  = "arXiv",
      primaryClass   = "astro-ph",
      SLACcitation   = "%%CITATION = ARXIV:0807.3551;%%"
}

@ARTICLE{Beutler_etal2011,
   author = {{Beutler}, F. and {Blake}, C. and {Colless}, M. and {Jones}, D.~H. and 
	{Staveley-Smith}, L. and others},
    title = "{The 6dF Galaxy Survey: baryon acoustic oscillations and the local Hubble constant}",
  journal = {MNRAS},
archivePrefix = "arXiv",
   eprint = {1106.3366},
     year = 2011,
    month = oct,
   volume = 416,
    pages = {3017-3032},
      doi = {10.1111/j.1365-2966.2011.19250.x},
   adsurl = {http://adsabs.harvard.edu/abs/2011MNRAS.416.3017B}
}

@ARTICLE{Anderson_BOSS2012,
       author = {{Anderson}, Lauren and  others},
        title = "{The clustering of galaxies in the SDSS-III Baryon Oscillation Spectroscopic Survey: baryon acoustic oscillations in the Data Release 9 spectroscopic galaxy sample}",
      journal = {MNRAS},
         year = 2012,
        month = dec,
       volume = {427},
       number = {4},
        pages = {3435-3467},
          doi = {10.1111/j.1365-2966.2012.22066.x},
archivePrefix = {arXiv},
       eprint = {1203.6594},
 primaryClass = {astro-ph.CO},
       adsurl = {https://ui.adsabs.harvard.edu/abs/2012MNRAS.427.3435A}
}

@ARTICLE{Kazin_etal2014,
   author = {{Kazin}, E.~A. and {Koda}, J. and {Blake}, C. and {Padmanabhan}, N. and 
	{Brough}, S. and others },
    title = "{The WiggleZ Dark Energy Survey: improved distance measurements to z = 1 with reconstruction of the baryonic acoustic feature}",
  journal = {MNRAS},
archivePrefix = "arXiv",
   eprint = {1401.0358},
     year = 2014,
    month = jul,
   volume = 441,
    pages = {3524-3542},
      doi = {10.1093/mnras/stu778},
   adsurl = {http://adsabs.harvard.edu/abs/2014MNRAS.441.3524K}
}

@ARTICLE{Alam_etal2017,
   author = {{Alam}, S. and {Ata}, M. and {Bailey}, S. and {Beutler}, F. and 
	{Bizyaev}, D. and others},
    title = "{The clustering of galaxies in the completed SDSS-III Baryon Oscillation Spectroscopic Survey: cosmological analysis of the DR12 galaxy sample}",
  journal = {MNRAS},
archivePrefix = "arXiv",
   eprint = {1607.03155},
     year = 2017,
    month = sep,
   volume = 470,
    pages = {2617-2652},
      doi = {10.1093/mnras/stx721},
   adsurl = {http://adsabs.harvard.edu/abs/2017MNRAS.470.2617A}
}

@ARTICLE{DES_Y1BAO2019,
       author = {{Abbott}, T.~M.~C. and {Abdalla}, F.~B. and {Alarcon}, A. and {Allam}, S. and {Andrade-Oliveira}, F. and others },
        title = "{Dark Energy Survey Year 1 Results: Measurement of the Baryon Acoustic Oscillation scale in the distribution of galaxies to redshift 1}",
      journal = {\mnras},
         year = 2019,
        month = mar,
       volume = {483},
       number = {4},
        pages = {4866-4883},
          doi = {10.1093/mnras/sty3351},
archivePrefix = {arXiv},
       eprint = {1712.06209},
 primaryClass = {astro-ph.CO},
       adsurl = {https://ui.adsabs.harvard.edu/abs/2019MNRAS.483.4866A}
}

@ARTICLE{DES_Y3BAO2022,
       author = {{Abbott}, T.~M.~C. and {Aguena}, M. and {Allam}, S. and {Amon}, A. and {Andrade-Oliveira}, F. and others },
        title = "{Dark Energy Survey Year 3 results: A 2.7\% measurement of baryon acoustic oscillation distance scale at redshift 0.835}",
      journal = {\prd},
         year = 2022,
        month = feb,
       volume = {105},
       number = {4},
          eid = {043512},
        pages = {043512},
          doi = {10.1103/PhysRevD.105.043512},
archivePrefix = {arXiv},
       eprint = {2107.04646},
 primaryClass = {astro-ph.CO},
       adsurl = {https://ui.adsabs.harvard.edu/abs/2022PhRvD.105d3512A}
}

@ARTICLE{DES_Y6BAO2024,
       author = {{Abbott}, T.~M.~C. and {Adamow}, M. and {Aguena}, M. and {Allam}, S. and {Alves}, O. and
others}, 
        title = "{Dark Energy Survey: A 2.1\% measurement of the angular baryonic acoustic oscillation scale at redshift zeff=0.85 from the final dataset}",
      journal = {\prd},
         year = 2024,
        month = sep,
       volume = {110},
       number = {6},
          eid = {063515},
        pages = {063515},
          doi = {10.1103/PhysRevD.110.063515},
archivePrefix = {arXiv},
       eprint = {2402.10696},
 primaryClass = {astro-ph.CO},
       adsurl = {https://ui.adsabs.harvard.edu/abs/2024PhRvD.110f3515A}
}

@article{eBOSS:2020yzd,
    author = {{Alam}, Shadab and {Aubert}, Marie and {Avila}, Santiago and {Balland}, Christophe and {Bautista}, Julian E. and others },
    collaboration = "eBOSS",
    title = "{Completed SDSS-IV extended Baryon Oscillation Spectroscopic Survey: Cosmological implications from two decades of spectroscopic surveys at the Apache Point Observatory}",
    eprint = "2007.08991",
    archivePrefix = "arXiv",
    primaryClass = "astro-ph.CO",
    doi = "10.1103/PhysRevD.103.083533",
    journal = "Phys. Rev. D",
    volume = "103",
    number = "8",
    pages = "083533",
    year = "2021"
}

@ARTICLE{Chan_DESY3BAO2022,
       author = {{Chan}, K.~C. and {Avila}, S. and {Carnero Rosell}, A. and {Ferrero}, I. and {Elvin-Poole}, J. and others },
        title = "{Dark Energy Survey Year 3 results: Measurement of the baryon acoustic oscillations with three-dimensional clustering}",
      journal = {\prd},
         year = 2022,
        month = dec,
       volume = {106},
       number = {12},
          eid = {123502},
        pages = {123502},
          doi = {10.1103/PhysRevD.106.123502},
archivePrefix = {arXiv},
       eprint = {2210.05057},
 primaryClass = {astro-ph.CO},
       adsurl = {https://ui.adsabs.harvard.edu/abs/2022PhRvD.106l3502C}
}

@ARTICLE{DESI_BAO2_2025,
       author = {{Abdul Karim}, M. and {Aguilar}, J. and {Ahlen}, S. and {Alam}, S. and {Allen}, L.  and others },
        title = "{DESI DR2 results. II. Measurements of baryon acoustic oscillations and cosmological constraints}",
      journal = {\prd},
         year = 2025,
        month = oct,
       volume = {112},
       number = {8},
          eid = {083515},
        pages = {083515},
          doi = {10.1103/tr6y-kpc6},
archivePrefix = {arXiv},
       eprint = {2503.14738},
 primaryClass = {astro-ph.CO},
       adsurl = {https://ui.adsabs.harvard.edu/abs/2025PhRvD.112h3515A}
}

@ARTICLE{DESI_BAO1_2024,
       author = {{Adame}, A.~G. and {Aguilar}, J. and {Ahlen}, S. and {Alam}, S. and {Alexander}, D.~M. and others},
        title = "{DESI 2024 VI: cosmological constraints from the measurements of baryon acoustic oscillations}",
      journal = {\jcap},
         year = 2025,
        month = feb,
       volume = {2025},
       number = {2},
          eid = {021},
        pages = {021},
          doi = {10.1088/1475-7516/2025/02/021},
archivePrefix = {arXiv},
       eprint = {2404.03002},
 primaryClass = {astro-ph.CO},
       adsurl = {https://ui.adsabs.harvard.edu/abs/2025JCAP...02..021A}
}

@article{Xu:2023vrl,
    author = "Xu, Kun and Jing, Y. P. and Zhao, Gong-Bo and Cuesta, Antonio J.",
    title = "{Evidence for baryon acoustic oscillations from galaxy\textendash{}ellipticity correlations}",
    eprint = "2306.09407",
    archivePrefix = "arXiv",
    primaryClass = "astro-ph.CO",
    doi = "10.1038/s41550-023-02035-4",
    journal = "Nature Astron.",
    volume = "7",
    number = "10",
    pages = "1259--1264",
    year = "2023"
}

@ARTICLE{ChisariDvorkin2013,
       author = {{Chisari}, Nora Elisa and {Dvorkin}, Cora},
        title = "{Cosmological information in the intrinsic alignments of luminous red galaxies}",
      journal = {\jcap},
         year = 2013,
        month = dec,
       volume = {2013},
       number = {12},
          eid = {029},
        pages = {029},
          doi = {10.1088/1475-7516/2013/12/029},
archivePrefix = {arXiv},
       eprint = {1308.5972},
 primaryClass = {astro-ph.CO},
       adsurl = {https://ui.adsabs.harvard.edu/abs/2013JCAP...12..029C}
}

@ARTICLE{OkumuraTaruya_2020,
       author = {{Okumura}, Teppei and {Taruya}, Atsushi},
        title = "{Anisotropies of galaxy ellipticity correlations in real and redshift space: angular dependence in linear tidal alignment model}",
      journal = {\mnras},
         year = 2020,
        month = mar,
       volume = {493},
       number = {1},
        pages = {L124-L128},
          doi = {10.1093/mnrasl/slaa024},
archivePrefix = {arXiv},
       eprint = {1912.04118},
 primaryClass = {astro-ph.CO},
       adsurl = {https://ui.adsabs.harvard.edu/abs/2020MNRAS.493L.124O}
}

@ARTICLE{DES_SNeBAO2025,
       author = {{DES Collaboration} and {Abbott}, T.~M.~C. and {Acevedo}, M. and {Adamow}, M. and {Aguena}, M. and others }, 
        title = "{Dark Energy Survey: implications for cosmological expansion models from the final DES Baryon Acoustic Oscillation and Supernova data}",
      journal = {arXiv e-prints},
         year = 2025,
        month = mar,
          eid = {arXiv:2503.06712},
        pages = {arXiv:2503.06712},
          doi = {10.48550/arXiv.2503.06712},
archivePrefix = {arXiv},
       eprint = {2503.06712},
 primaryClass = {astro-ph.CO},
       adsurl = {https://ui.adsabs.harvard.edu/abs/2025arXiv250306712D}
}

@ARTICLE{DES_SNeY5_2024,
       author = {{DES Collaboration} and {Abbott}, T.~M.~C. and {Acevedo}, M. and {Aguena}, M. and {Alarcon}, A. and others },
        title = "{The Dark Energy Survey: Cosmology Results with {\ensuremath{\sim}}1500 New High-redshift Type Ia Supernovae Using the Full 5 yr Data Set}",
      journal = {\apjl},
         year = 2024,
        month = sep,
       volume = {973},
       number = {1},
          eid = {L14},
        pages = {L14},
          doi = {10.3847/2041-8213/ad6f9f},
archivePrefix = {arXiv},
       eprint = {2401.02929},
 primaryClass = {astro-ph.CO},
       adsurl = {https://ui.adsabs.harvard.edu/abs/2024ApJ...973L..14D}
}

@ARTICLE{Lamman_IAguide2024,
       author = {{Lamman}, Claire and {Tsaprazi}, Eleni and {Shi}, Jingjing and {{\v{S}}ar{\v{c}}evi{\'c}}, Nikolina Niko and  others},
        title = "{The IA Guide: A Breakdown of Intrinsic Alignment Formalisms}",
      journal = {The Open Journal of Astrophysics},
         year = 2024,
        month = feb,
       volume = {7},
          eid = {14},
        pages = {14},
          doi = {10.21105/astro.2309.08605},
archivePrefix = {arXiv},
       eprint = {2309.08605},
 primaryClass = {astro-ph.CO},
       adsurl = {https://ui.adsabs.harvard.edu/abs/2024OJAp....7E..14L}
}

@ARTICLE{Chisari2025,
       author = {{Chisari}, Nora Elisa},
        title = "{A rising tide: intrinsic alignments since the turn of the millennium}",
      journal = {The Astronomy and Astrophysics Review},
         year = 2025,
        month = oct,
       volume = {33},
       number = {1},
          eid = {5},
        pages = {5},
          doi = {10.1007/s00159-025-00161-8},
archivePrefix = {arXiv},
       eprint = {2510.15738},
 primaryClass = {astro-ph.CO},
       adsurl = {https://ui.adsabs.harvard.edu/abs/2025A&ARv..33....5C}
}

@ARTICLE{Joachimi_alignment2015,
       author = {{Joachimi}, Benjamin and {Cacciato}, Marcello and {Kitching}, Thomas D. and {Leonard}, Adrienne and {Mandelbaum}, Rachel and others },
        title = "{Galaxy Alignments: An Overview}",
      journal = {\ssr},
         year = 2015,
        month = nov,
       volume = {193},
       number = {1-4},
        pages = {1-65},
          doi = {10.1007/s11214-015-0177-4},
archivePrefix = {arXiv},
       eprint = {1504.05456},
 primaryClass = {astro-ph.GA},
       adsurl = {https://ui.adsabs.harvard.edu/abs/2015SSRv..193....1J}
}

@ARTICLE{TroxelIshak2015,
       author = {{Troxel}, M.~A. and {Ishak}, Mustapha},
        title = "{The intrinsic alignment of galaxies and its impact on weak gravitational lensing in an era of precision cosmology}",
      journal = {\physrep},
         year = 2015,
        month = feb,
       volume = {558},
        pages = {1-59},
          doi = {10.1016/j.physrep.2014.11.001},
archivePrefix = {arXiv},
       eprint = {1407.6990},
 primaryClass = {astro-ph.CO},
       adsurl = {https://ui.adsabs.harvard.edu/abs/2015PhR...558....1T}
}

@ARTICLE{Friedrich_etal2021,
       author = {{Friedrich}, O. and {Andrade-Oliveira}, F. and {Camacho}, H. and {Alves}, O. and {Rosenfeld}, R. and others },
        title = "{Dark Energy Survey year 3 results: covariance modelling and its impact on parameter estimation and quality of fit}",
      journal = {\mnras},
         year = 2021,
        month = dec,
       volume = {508},
       number = {3},
        pages = {3125-3165},
          doi = {10.1093/mnras/stab2384},
archivePrefix = {arXiv},
       eprint = {2012.08568},
 primaryClass = {astro-ph.CO},
       adsurl = {https://ui.adsabs.harvard.edu/abs/2021MNRAS.508.3125F}
}

@article{DNF,
	adsurl = {http://adsabs.harvard.edu/abs/2016MNRAS.459.3078D},
	archiveprefix = {arXiv},
	author = {{De Vicente}, J. and {S{\'a}nchez}, E. and {Sevilla-Noarbe}, I.},
	doi = {10.1093/mnras/stw857},
	eprint = {1511.07623},
	journal = {\mnras},
	month = jul,
	pages = {3078-3088},
	title = {{DNF - Galaxy photometric redshift by Directional Neighbourhood Fitting}},
	volume = 459,
	year = 2016}

\end{document}